\documentclass{article}
\usepackage[utf8]{inputenc}
\usepackage{authblk}
\usepackage{graphicx,mathtools}

\title{Literature Review of the Pioneering Approaches
in Cloud-based Search Engines Powered by LETOR Techniques}
\author[1]{Gizem Gezici}
\affil[1]{Sabanci University, Istanbul, Turkey}
\affil[2]{Huawei Turkey R\&D Center, Istanbul, Turkey}
\date{}

\begin{document}

\maketitle

\begin{abstract}
Search engines play an essential role in our daily lives. Nonetheless, they are also very crucial in enterprise domain to access documents from various information sources. Since traditional search systems index the documents mainly by looking at the frequency of the occurring words in these documents, they are barely able to support natural language search, but rather keyword search.
It seems that keyword based search will not be sufficient for enterprise data which is growing extremely fast. Thus, enterprise search becomes increasingly critical in corporate domain. In this
report, we present an overview of the state-of-the-art technologies in literature for three main purposes: i) to increase the retrieval performance of a search engine, ii) to deploy a search platform
to a cloud environment, and iii) to select the best terms in expanding queries for achieving even a higher retrieval performance as well as to provide good query suggestions to its users for a better user experience.
\end{abstract}

\section{Three Main Project Modules}
The overall architecture of our project is composed of three fundamental modules, namely~\emph{LEarning TO Rank (LETOR)}, ~\emph{Cloud Search}, and ~\emph{Query Expansion \& Suggestion}. Note that the overall system was inspired by the Yahoo search team article that comprehensively describes not only successes but also struggles with possible
solutions in increasing Yahoo search engine's retrieval performance. The article is well-written best paper award-winner that is a good reference to understand all aspects of an end-to-end search engine. Hence, we established an overall structure of our search system using this paper as a reference~\cite{yin2016ranking}. One can find the project module details with the referenced papers that are elaborately discussed below.

\subsection{LEarning TO Rank (LETOR)}
The first module of the overall system utilizes and improves the state-of-the-art solutions in the literature to establish a successful search platform. In IR, the performance of a search engine is evaluated by measuring the retrieval performance i.e.,
retrieving most relevant documents at higher positions in search results. It has been achieved by employing traditional machine learning approaches and recently, also deep learning-based techniques has become part of this research area that is called
LETOR.

In the scope of the LETOR module, we explore the feature space by looking for the most useful features in document retrieval task. In our feature set, while some of the features exploit the textual
content, others may convey the click information in order to represent the given dataset (training data) well enough for better learning.

In the~\emph{Elasticsearch Plug-in Based Ranking} submodule, we explored textual features whereas in the ~\emph{Click Graph Based Ranking}, we delved into click-through analysis and integrated suitable features for that analysis to our feature set. In addition to these two submodules in ~\emph{Deep Learning Based Ranking}, it seems that we will not need to fullfill any feature engineering work, rather we allow the end-to-end deep learning system to extract features and learn model from the training dataset simultaneously. More details about the submodules can be found in the corresponding sections below. Furthermore in the ~\emph{Click Graph Based Ranking} which is the last submodule of LETOR, our main aim is to learn the semantic vector representations of query-document pairs without using deep learning techniques, rather a graph approach on the training dataset.

\subsubsection{Dataset}
The training dataset is composed of query-document pairs with click information as well as the relevance grades of the documents for the
given queries. For each query-document pair, we have query, document title and click information. Moreover, there exists document url information that can be used (if needed) to fetch the whole document from the website for further analysis i.e., analysing snippet, keywords (if any) or important sections in the document.

In addition to these, to evaluate our search platform we need to have relevance grades of the documents for their query pairs in the dataset. However, relevance information (ground truth) may not be available all the time and in that case we would encounter a very common problem of supervised approaches, namely the difficulty of having labelled data. Manual labelling is a costly process, therefore we may need alternative approaches such as transforming click information into relevance labels as in~\cite{joachims2002optimizing, carterette2007evaluating}. The detailed step-by-step transformation procedure can be found in our second-term progress report.

\subsubsection{Elasticsearch Plug-in Based Ranking}
In this module, initially we obtain Elasticsearch~\cite{gormley2015elasticsearch} features about query-document pairs in the click-logs. On top of this, we compute rather more complex features that are commonly used in IR systems such as tf, matched terms count, BM25, and related features that show the similarity of queries and corresponding documents in the logs. With the help of these additional features, our aim is to
exploit query-document pairs more, in this way we can obtain more information and feed the system with this enriched input for training. Hopefully, the system will be able to learn better in comparison to only Elasticsearch features case, from the click-logs (training data) and a higher accuracy for document retrieval task will be achieved.

The procedure of this submodule is as follows, each query-document pair is initially denoted as a feature vector and model is learned from the feature file by using a ranking algorithm from RankLib~\cite{dang2013lemur}. In this part of the overall system, our contribution is to propose more complex text-similarity features in addition to the relatively simple Elasticsearch features, hoping to increase the retrieval performance of our search platform.

\subsubsection{Deep Learning-Based Ranking}
Deep learning methods have become quite popular in recent years. Researchers have obtained the best accuracies so far with deep learning models in various research areas such as computer vision and NLP. Some of these computer vision and NLP tasks, on which DL-based approaches outperform traditional methods, are object detection, word embeddings, part-of-speech-tagging, finding dependencies in a given sentence as well as relatively more complex tasks like sentiment analysis and image captioning. Owing to the big achievements of deep learning on a broad range of problems, big technology companies like Google, Facebook, Amazon and Yahoo has invested in it continuously.

Additionally, deep learning methods also have been applied to increase the retrieval performance of search engines. Yahoo~\cite{yin2016ranking} and Microsoft~\cite{shen2014latent} established more successful search platforms with deep learning models in comparison to state-of-the traditional approaches. In these published works, semantic queries are supported instead of simple keyword matching by taking into account of query context, along with user intent. You can find a list of pioneering studies that employ deep learning approaches to build a search engine with high retrieval performance in more detail below. Note that all of these works utilize supervised deep learning models and in this framework, click-logs is the training set in which queries are inputs and corresponding clicked documents are seen as outputs to train the chosen deep learning architecture.

\paragraph{Paper Review: Learning Semantic Representations Using Convolutional Neural Networks for Web Search~\cite{shen2014learning}}

\paragraph{Motivation: }
Improving the modeling contextual information in click-log queries/documents and capturing it in a fine-grained manner.

\paragraph{Method: }
The paper proposes a series of new latent semantic models based on a ~\emph{convolutional neural network} to learn semantic word embeddings for search queries and Web documents. Initially, local contextual information at the word n-gram level is modeled by applying the convolution-max pooling operation. Subsequently, in order to constitute a global feature vector, salient local features in a word sequence are combined. As a final step, the high-level semantic feature vector of the input word sequence is extracted to form a global vector representation. To train the architecture, the proposed models are trained on click-through data by maximizing the conditional likelihood of clicked documents given a query, applying stochastic gradient ascent. \\

The closest work is DSSM~\cite{huang2013learning}, which is declared to outperform significantly semantic hashing and other traditional semantic models. Compared with DSSM, C-DSSM has a convolutional layer in which each word is projected within a context window to a local contextual feature vector.

\begin{figure}[!t]
    \centering
    {\includegraphics[scale = 0.85]{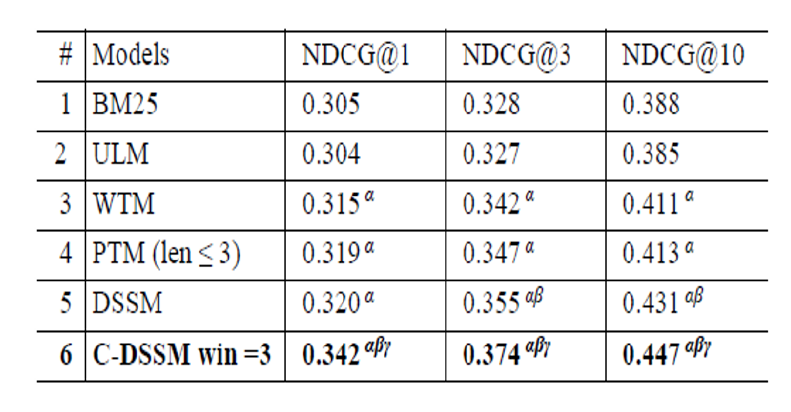}}
    \caption{Superscripts $\alpha, \beta$, and $\gamma$ indicate 
    statistically significant improvements ($p < 0.05$) over $BM25$, $PTM$, and $DSSM$, respectively.}
    \label{fig:table1}
\end{figure}

\paragraph{Experimental Results: }
The retrieval model has been evaluated on a large-scale real world data set that contains 12,071 English queries sampled from one-year period of query log files. The evaluation metric is Normalized Discounted Cumulative Gain (NDCG) and only document titles were
used for ranking. In the experiments, the click-through data used in training include 30 million of query/clicked document title pairs. \\

The proposed C-DSSM was compared with a set of baseline models, including BM25, the unigram language model (ULM), phrase-based translation model (PTR), word-based translation model (WTM), and the closest work to the current architecture, DSSM. As shown in  Figure~\ref{fig:table1}, the C-DSSM outperforms all the state-of-the-art approaches with a significant margin.

\paragraph{Paper Review: A Latent Semantic Model with Convolutional-Pooling Structure for Information Retrieval~\cite{shen2014latent}}

\paragraph{Motivation: }
In spite of the notable achievements obtained in recent studies, still all the prior latent semantic models treat a query (or a document) as a BoWs. Therefore, they are not effective in detecting contextual structures of a query (or a document).

\begin{figure}[!h]
    \centering
    \includegraphics[scale = 0.85]{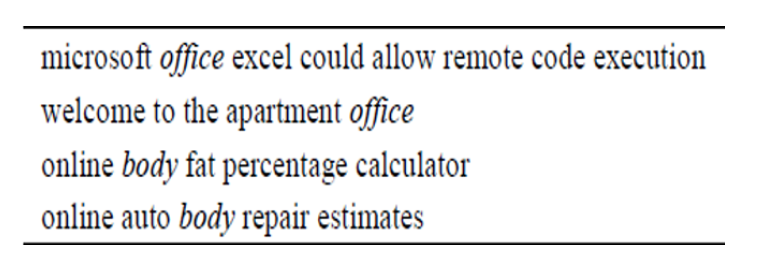}
    \caption{Sample document titles. The text is lower-cased and punctuation removed. The same word, e.g., "office", has different meanings depending on its contexts.}
    \label{fig:fig1}
\end{figure}

As shown in Figure~\ref{fig:fig1} with several examples of document titles, the contextual information is very valuable in the task of semantic search and without this information, it seems that system fails to achieve high retrieval performance.

\begin{figure}[!h]
    \centering
    \includegraphics[scale = 0.85]{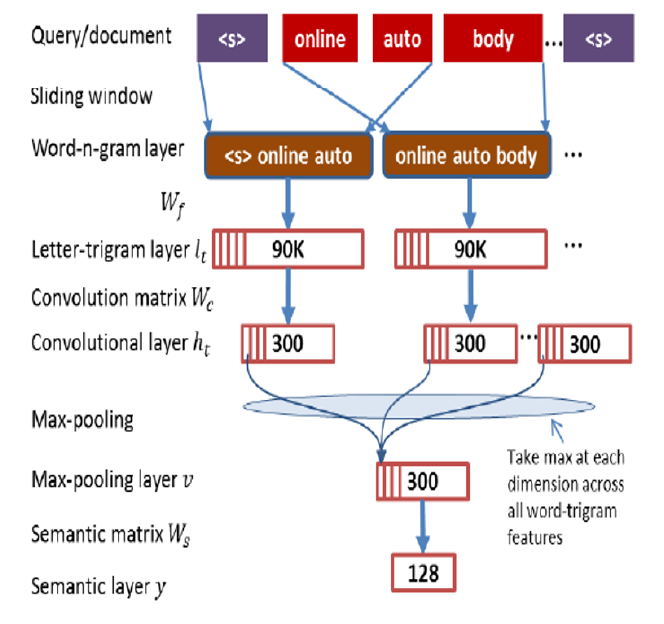}
    \caption{The CLSM maps a variable-length word sequence to a low-dimensional vector in a latent semantic space. A word contextual window size (i.e. the receptive field) of three is used in the illustration. Convolution over word sequence via learned matrix $W_c$ performed implicitly via the earlier layer's mapping with a local receptive field. The dimensionalities of the convolutional layer and the semantic layer are set to 300 and 128 in the illustration, respectively. The max operation across the sequence is applied for each of 300 feature dimensions separately. (Only the first dimension is shown to avoid figure clutter.)}
    \label{fig:fig3}
\end{figure}

\begin{figure}[!h]
    \centering
    \includegraphics[scale = 0.85]{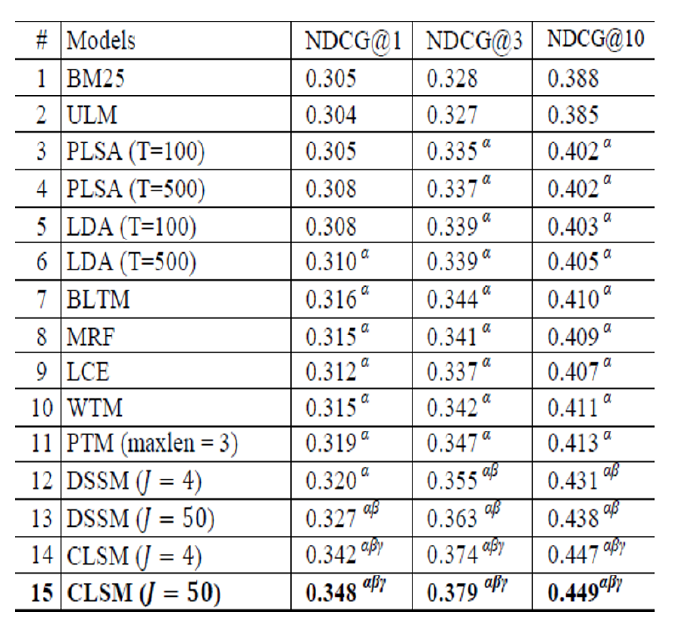}
    \caption{Comparative results with the previous state of the art approaches. BLTM, WTM, PTM, DSSM, and CLSM use the same click-through data for learning. Superscripts $\alpha$, $\beta$, and $\gamma$ indicate statistically significant improvements ($p < 0.05$) over $BM25$, $PTM$, and $DSSM (J = 50)$, respectively. (Models of \#1, \#2, \#11, \#12, and \#13 have been also used in the previous C-DSSM paper for comparison.)}
    \label{fig:fig4}
\end{figure}

\paragraph{Method: }
In this study, a new latent semantic model that incorporates a convolutional-pooling structure over word sequences to learn low-dimensional, semantic vector representations for search queries/documents in the click-log. In order to detect the rich contextual structures, the procedure starts with each word within a sliding window (temporal context window) in a word sequence to directly capture contextual features at the word n-gram level. \\

In order to use the CLSM for IR, given a query and corresponding Web documents to be ranked, firstly the semantic vector representations for the query and all the documents using the architecture as described above. Then, a semantic relevance score is computed by measuring the cosine similarity between the semantic vectors of the query
Q and each document D in the click-log which is used for training. In this work, the underlying assumption is that a query is relevant to the documents that are clicked on for that query, and train the CLSM on the click-through data accordingly. The high-level C-DSSM architecture is depicted in Figure~\ref{fig:fig3}.

\paragraph{Experimental Results: }
The evaluation is done on a Web document ranking task using
a large-scale, real-world data set that contains. 12,071 English queries sampled from one-year query log files. Each query-document pair has a relevance label manually annotated on a 5-level relevance scale: $bad$, $fair$, $good$, $excellent$, and $perfect$, corresponding to 0 ($bad$) to 4 ($perfect$). \\

Results demonstrate that in retrieval performance, the proposed model significantly outperforms other state-of-the-art semantic models, which were prior to this work. BM25 and ULM are used as baselines and both use term vector representation. PLSA~\cite{hofmann1999probabilistic} was trained on documents only using MAP estimation with different number of topics, T. BLTM is the best performer bilingual topic model in~\cite{gao2011clickthrough}. MRF models the term dependency proposed in~\cite{metzler2005markov}. LCE is a latent concept expansion model as described in~\cite{metzler2007latent} that leverages the term-dependent information. WTM, a word-based translation model and PTM, phrase-based translation model were implemented as described in~\cite{gao2010clickthrough}. Lastly DSSM, which is the best variant of DSSM proposed in~\cite{huang2013learning}, is used for comparison by changing the number of negative samples J. Overall comparative results can be found in Figure~\ref{fig:fig4}.

\begin{figure}[!h]
    \centering
    \includegraphics[scale = 0.85]{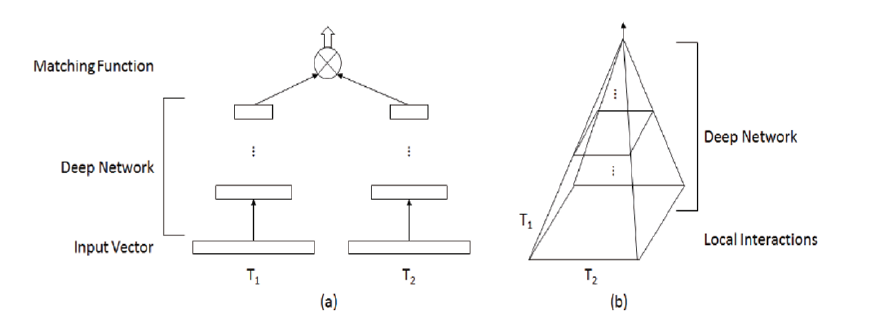}
    \caption{Two types of deep matching models: (a) Representation-focused models employ a Siamese (symmetric) architecture over the text inputs; (b) Interaction-focused models employ a hierarchical deep architecture over the local interaction matrix.}
    \label{fig:fig5}
\end{figure}

\paragraph{Paper Review: A Deep Relevance Matching Model for Ad-hoc Retrieval~\cite{guo2016deep}}

\paragraph{Motivation: }
Although in recent years, deep neural networks have achieved successful results in distinct research areas such as computer vision, and natural language processing (NLP) tasks as mentioned above, few positive results have been
reported in ad-hoc retrieval tasks. One of the main reasons behind this may be stemmed from the fact that many essential characteristics of the ad-hoc retrieval task have not yet been well addressed in deep models. The ad-hoc retrieval task is commonly formalized as a matching problem between textual contents of query and document in prior works using deep learning, and viewed in a similar way to many NLP asks such as paraphrase identification, question answering and automatic conversation. \\

However, researchers defend that the ad-hoc retrieval task is essentially about relevance matching while most NLP matching tasks solve semantic matching problem, and there exist some major differences between these matching tasks. Achieving high performance in relevance matching requires
proper handling of the~\textbf{exact matching signals, query term importance and diverse matching requirements}.

\paragraph{Method: }
In this paper, a novel deep relevance matching model (DRMM) for ad-hoc retrieval has been proposed. Specifically, in the proposed model, a joint deep architecture at the query term level is employed for relevance matching. By applying matching histogram mapping, a feed forward matching network, and a term gating network, the researchers can effectively incorporate the three relevance matching factors mentioned above. \\

So far, to solve the matching problem and treat ad-hoc retrieval as an NLP task, different deep matching models have been presented. These
deep learning architectures can be mainly categorized into two types based on their model architecture: a) representation-focused models, and b) interaction-focused models as depicted in Figure~\ref{fig:fig5}. \\

The first type of models, the representation-focused model, tries to build a good representation for a single text with a deep neural network, and then carries out matching between the
compositional and abstract text representations. For instance, DSSM~\cite{huang2013learning} and C-DSSM~\cite{shen2014learning} can be categorized as the representation-focused models. The other is the interaction-focused model and in this type of model, local interactions between two pieces of text are formed, and then uses deep neural networks to detect hierarchical interaction patterns for matching. Deep Match~\cite{guo2016deep} is an example for interaction-focused models and the Deep Relevance Matching Model (DRMM) proposed in this work can also be put in this category. \\

In this study, the underlying hypothesis is that semantic matching and relevance matching are not the same thing. In fact, they are quite different problems and specifically, researchers point out three fundamental differences between these two concepts. In many NLP tasks such as paraphrase
identification, question answering and automatic conversation, the matching is mainly related to~\emph{semantic matching}, i.e. identifying the semantic meaning and inferring the semantic relations between two pieces of text. The matching in ad-hoc retrieval task, on the other hand, is essentially about ~\emph{relevance matching}, i.e. identifying whether a document is relevant to a given query. These two different matching problems emphasize three distinct elements to find out solutions for NLP (semantic matching) and ad-hoc retrieval (relevance matching) respectively. In discussing these three
factors, our aim is to show why we need to differentiate ~\emph{relevance matching} from~\emph{semantic matching}.

\begin{enumerate}
\item \textbf{Similarity matching signals vs. Exact
matching signals:} In our ad-hoc retrieval task, although more complex metrics have also been proposed, the exact matching of
query terms in documents is still the most important signal. Unlikely, NLP tasks need to detect semantically related words which can convey the same meaning even if they do not share any common word or phrases. This also clarifies why some traditional IR models, which are simply based on exact matching, e.g., BM25, can work fairly well for ad-hoc
retrieval while other traditional NLP models cannot show a similar performance for NLP-related tasks.

\item \textbf{Compositional meanings vs. Query
term importance:} In the scope of NLP, it is useful to extract grammatical structures to capture compositional meaning rather than seeing sentences as a BoWs in the given text. On the other hand, in ad-hoc retrieval, queries are composed of mainly short
and keyword based phrases without complex grammatical structures. Therefore, in our case it is crucial to take into account of term importance instead of grammatical structures.

\item \textbf{Global matching requirements vs. Diverse matching requirements:} In the literature, there are various hypotheses about document length such as Verbosity Hypothesis and Scope Hypothesis. The Verbosity Hypothesis follows an assumption that a long document covers similar content but with more words. Conversely, the Scope Hypothesis considers a long document is composed of a number of unrelated appended short documents. Hence, in terms of the Verbosity Hypothesis, relevance matching might be global for the assumption that short documents have a concentrated topic, whereas based on the Scope Hypothesis, partial relevance, the relevance of different parts of the document to a query is necessary. On the other hand, semantic matching mostly requires global matching with the aim of inferring the semantic relations from the whole text. \\

Based on these important differences between relevance matching in ad-hoc retrieval and semantic matching in many NLP tasks, it is clear that we need to establish a deep model architecture which incorporates these differences into the model properly. Herein, previously proposed architectures seem to be deficient; thus a novel deep learning architecture specifically designed for relevance matching in ad-hoc retrieval, namely deep relevance matching model (DRRM), has been suggested. Note that the introduced architecture is akin to interaction-focused rather than representation-focused models since detailed matching signals are very crucial and they are inevitably lost in the latter group of models.

\end{enumerate}

\begin{figure}[!t]
    \centering
    \includegraphics[scale = 0.80]{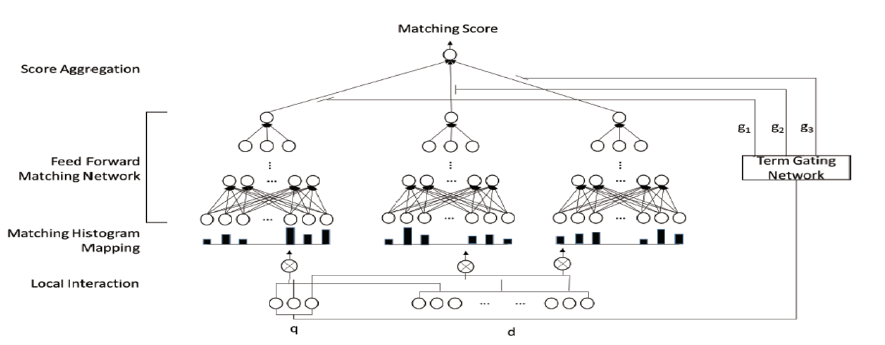}
    \caption{Architecture of DRMM}
    \label{fig:fig6}
\end{figure}

\underline{The Proposed Architecture:} 
The introduced model applies a joint deep architecture at the
query term level over the local interactions between query and document terms for relevance matching. In this way, query term importance can be modelled and in the following steps, the
contribution of each query term to the relevance score can be measured, hence different weights can be assigned to these terms, accordingly. Note that employing a joint deep architecture at the query term level is one important difference of the introduced model from existing interaction-focused models.

Initially, based on term embeddings, local interactions between each query-document pair of terms are established. Then, for each query term, the variable-length local interactions are transformed into a fixed-length matching histogram. As a subsequent step based on the matching histogram, a feed forward matching network is employed to learn hierarchical matching patterns and a matching score is generated for each query term.
Finally, the overall matching score is produced by aggregating the scores from each single query term with a term gating network in which the aggregation weights are computed. The proposed model architecture, DRMM is depicted in Figure~\ref{fig:fig6}.

\paragraph{Experimental Results: }
In this work, given the limited number of queries for each collection, 5-fold cross validation is conducted to avoid
over-fitting. Mean Average Precision (MAP) is used for parameter optimization. For evaluation, the top-ranked 1,000 documents are compared using the three commonly used IR metrics, namely MAP, normalized discounted cumulative gain at rank 20 (nDCG@$20$), and precision at rank 20 (P@$20$). Statistical differences between state-of-
the-art models are computed using the Fisher randomization test~\cite{smucker2007comparison}.\\

\begin{figure}[!t]
    \centering
    \includegraphics[scale = 0.83]{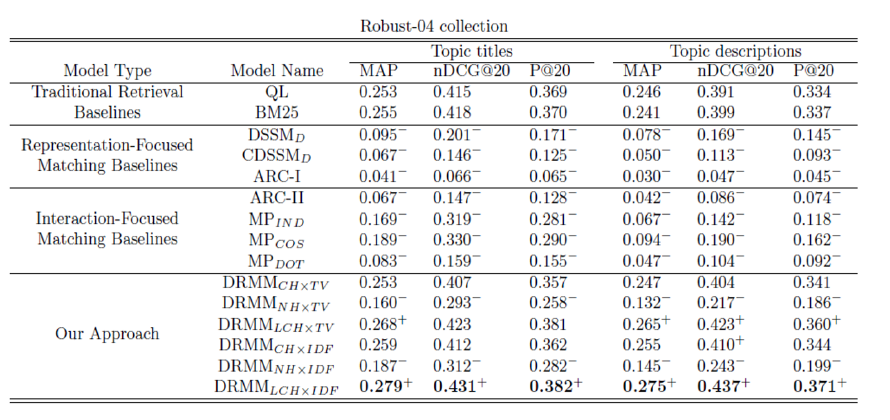}
    \caption{Comparison of different retrieval models over the Robust-04 collection. Significant improvement or degradation wrt QL is indicated (+/-) ($p-value < 0.05$).}
    \label{fig:fig7}
\end{figure}

In the evaluation part, two TREC collections are used for evaluation. However, we report the experimental results only for one of these datasets, namely Robust-04 collection in this report. In the scope of traditional retrieval baselines, we already mentioned BM25 and QL refers to query likelihood model based on Dirichlet smoothing~\cite{zhai2004study} which is one of the best performing language models. Results are displayed in Table~\ref{fig:fig7}. It can be concluded that all the representation-focused models perform significantly worse than the traditional retrieval models which demonstrates the unsuitability of these models for relevance matching. This is a very striking result that supports the paper's claim on the relevance/semantic matching difference. Please refer to the paper itself for comprehensive results and the related discussion.

\begin{figure}[!t]
    \centering
    \includegraphics[scale = 0.83]{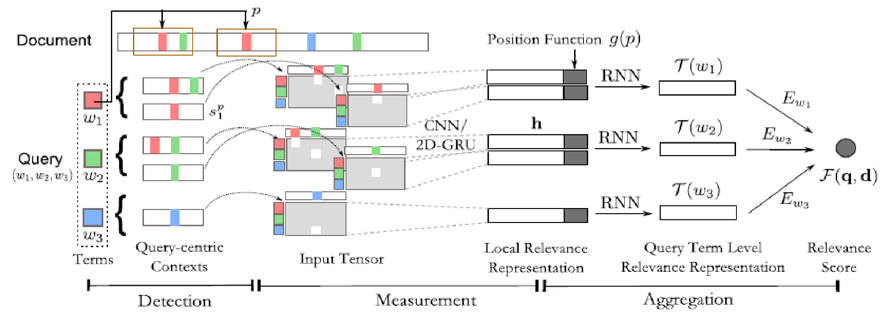}
    \caption{Architecture of DeepRank}
    \label{fig:fig8}
\end{figure}

\begin{figure}[!t]
    \centering
    \includegraphics[scale = 0.83]{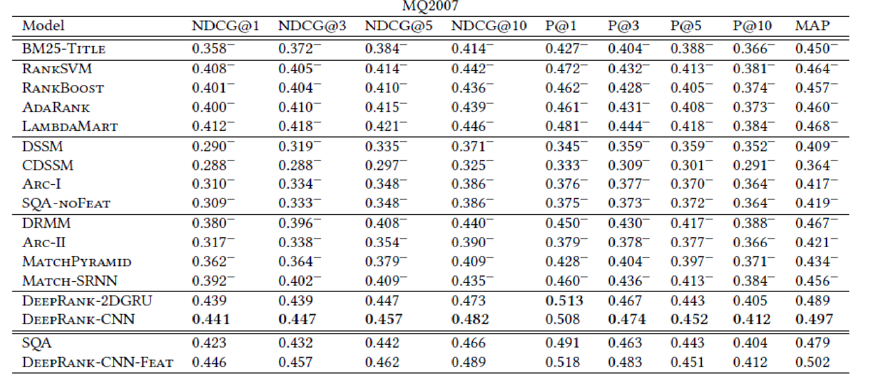}
    \caption{Performance comparison of different models on MQ2007. Significant improvement or degradation wrt DeepRank-CNN is denoted as (-) ($p-value < 0.05$).}
    \label{fig:fig9}
\end{figure}

\paragraph{Paper Review: DeepRank: A New Deep Architecture for Relevance Ranking in Information Retrieval~\cite{pang2017deeprank}}

\paragraph{Motivation: }
Existing deep IR models such as DSSM~\cite{huang2013learning} and C-DSSM~\cite{shen2014learning} generate ranking scores
by directly applying neural networks, without a proper understanding of the relevance (i.e. differentiating semantic matching and relevance matching. Although the previous architecture, DRMM semantic matching has capability of distinguishing these two different matching problems, it does not explicitly model the relevance generation process and fails to capture important IR features such as passage retrieval intrinsic and proximity
heuristics. \\

\paragraph{Method: }
The proposed model aims to mimic relevance label generation steps applied by the
human judgement process. According to this process, the relevance label generation comprises three steps: i) relevant locations are determined, ii) local relevances are determined, iii) local relevances are aggregated to output the relevance label. Initially, to extract the relevant contexts, a detection procedure is devised. Then, to detect the local relevances by using a convolutional neural network (CNN) or two-dimensional gated recurrent units (2D-GRU) as a measure network. Finally, an aggregation network with sequential
integration and term gating mechanism is applied to produce a global relevance score. \\

Note that DeepRank as illustrated in Figure~\ref{fig:fig8}, was proposed by the same lab to alleviate the weaknesses of their previous model, DRMM. It seems that DeepRank well captures significant IR (relevance matching) characteristics that distinguish relevance matching from semantic matching, including exact/semantic matching signals, proximity heuristics, query term importance, and diverse relevance requirement.

\paragraph{Experimental Results: }
Extensive experiments are conducted to evaluate DeepRank against state-of-the-art models such as learning to rank methods, and existing deep learning models.
For evaluation NDCG, Precision, and MAP metrics are used. \\

Experiments show that LETOR4.0 (MQ2007, MQ2008) benchmark and a large scale click-through data show that DeepRank can significantly outperform all the baseline methods. More specifically, in making comparison to learning to
rank methods, DeepRank performs even better than these models, whereas other existing deep learning methods show much worse performance. Comparison results on one of the LETOR4.0 benchmark dataset, MQ2007 are displayed in Table~\ref{fig:fig9}. For comprehensive results, please refer to the evaluation part of the paper.

\subsubsection{Click Graph Based Ranking}
For LTR, in our last submodule we utilize click graph idea to obtain more and different type of information (if feasible) from the click-through logs. Note that in this submodule, we only refer to one main paper since it is the most recent and successful work in this area.

\paragraph{Paper Review: Learning Query and Document Relevance from a Web-scale Click Graph~\cite{jiang2016learning}}

\paragraph{Motivation: }
Click-through logs contain rich and valuable information. However, the click information is sometimes noisy and its coverage is limited since there is a huge number of all possible relevant query-document pairs which leads to sparsity for the click-based features. The sparsity and noise problems affect the overall click-based feature quality negatively, especially for less popular (e.g. tail queries) queries. To overcome these problems, an effective way is to use click
and content information simultaneously. For this reason, learning a vector representation for both queries and documents in the same semantic space is needed. \\

Previous state-of-the-art approaches represent queries/documents in the same space such as traditional methods, which learn low-rank vectors, or direct text matching methods like BM25 and the language models. However, prior methods have its own advantages, it has also some weaknesses. For instance, low rank embedding hurts interpretability and debuggability of the ranking function because individual dimension in the latent space is hard to interpret and direct text matching methods suffer from the lexical gap between queries and documents. Moreover, we need an approach for click-absent queries, i.e. queries which have never been observed in the search logs.

\begin{figure}[!t]
    \centering
    \includegraphics[scale = 0.83]{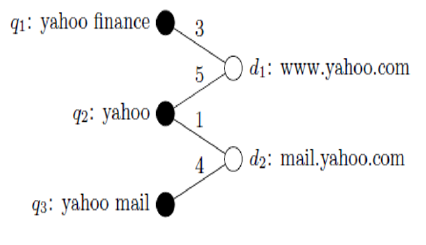}
    \caption{An example of click-through bipartite graph}
    \label{fig:fig10}
\end{figure}

\paragraph{Method: }
To overcome all these challenges, the
paper proposes a propagation approach to learn
vector representation by using both content and
click information. These vector representations
can directly improve the retrieval performance for
queries and documents that exist in the click logs,
i.e. click-existing queries. A sample click graph is
shown in Figure~\ref{fig:fig10}. \\

Furthermore, for click-absent queries and documents, a two-step vector estimation algorithm is proposed which utilizes partial information of the vectors in the bipartite graph that is already created by propagation. In this way, researchers
aim to significantly improve the coverage of the vectors, which is specifically critical for long-tail queries in web-search, i.e. the queries containing
keywords that are more specific and less common than other keywords.

\begin{figure}[!t]
    \centering
    \includegraphics[scale = 0.83]{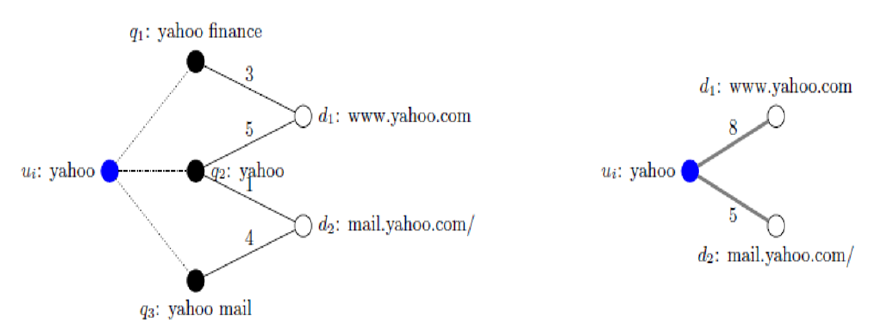}
    \caption{An example of unit vector generation. (The black thin lines are the edge of the click
graph, while the edges represented by the gray thick lines indicate the pseudo clicks between units
and documents.)}
    \label{fig:fig11}
\end{figure}

\paragraph{Vector Propagation Algorithm: }
The goal of this propagation algorithm is to learn the vector representation of queries and documents in the same semantic space (either the query or the document space). The algorithm starts from one side (query or document) and the vectors (query or document) are initialized with the content information and the vectors are propagated to their corresponding connected nodes on the side of the click graph. \\

Using bag-of-words model to generate the vector representations using query words to represent queries, and document titles to represent documents is a more intuitive way for vector generation. However, this procedure leads to the lexical gap between queries and documents. Thus, in the paper, researchers prefer to represent queries in query semantic space and documents in document semantic space. Note that co-clicks between queries and documents show the importance level
of the propagating terms by weighting the vector representations. In this way, significant terms become more prominent while less informative terms are eventually filtered out. \\

Apart from these, in order to incorporate click-absent queries to the already established graph, the paper uses a unit vector generation algorithm that is shown in Figure~\ref{fig:fig11}. In this procedure, firstly queries and documents titles are broken down into different units (e.g., ngrams) and vector representations for each unit are learned by using the vectors that have been already learned from the click-existing queries/documents in the graph. Then, a weight score is learned
for each unit by a regression model, and finally the vectors are estimated for the click-absent queries/documents by a linear combination of the unit vectors. Owing to this procedure, the click-absent queries/documents will be connected to the click graph through the unit vectors by utilizing the partial information in the already established click graph, even if the complete version of queries do not exist in the click-logs. In this way, high-quality representations are generated also for click-absent query/documents which substantially affects the retrieval performance of search engines in practice.

\begin{figure}[!t]
    \centering
    \includegraphics{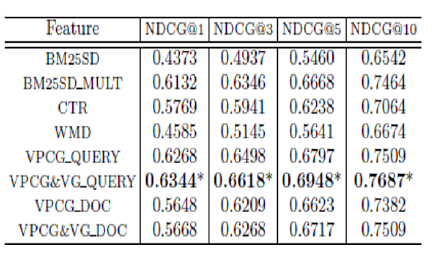}
    \caption{As an individual ranking model (Two-tailed t-test is done for paired data where each pair is VPCG \& VG QUERY and any of the other methods, and * indicates $p-value < 0.01$ for all tests.))}
    \label{fig:fig12}
\end{figure}

\begin{figure}[!t]
    \centering
    \includegraphics{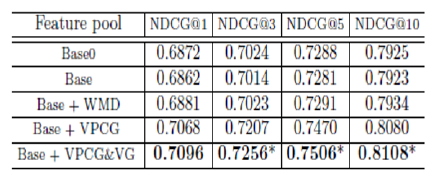}
    \caption{As a feature (Two-tailed t-test is done for paired data where each pair is "Base + VPCG \& VG" and any of the other methods, and
* indicates $p-value < 0.01$ for all tests.)))}
    \label{fig:fig13}
\end{figure}

\paragraph{Experimental Results: }
Researchers establish the click-through from a major commercial search engine's search log. There are approximately 25 billion co-clicked query-document pairs,
containing about 8 billion unique queries and 3 billion unique documents. This dataset was used as training set to build the graph. Then, for evaluation (i.e. for investigating if the relevance score learned by the proposed algorithm can help to improve ranking in a learning-to-rank framework), another dataset was used which is composed of 63k queries and 775k query-document pairs as training instances and 16k queries with 243k query-document pairs as test set. \\

The relevance score of each pair (``perfect", ``excellent", ``good", ``fair", ``bad") is annotated by human annotators. The evaluation is done in two distinct ways: i) the learned relevance score can be either used directly to rank documents, or ii) added to the feature vector in a learning-to-rank framework. You can find the results in Table~\ref{fig:fig12} and~\ref{fig:fig13} below. Results show that the proposed method helps to improve ranking results in both cases.

\begin{figure}[!t]
    \centering
    \includegraphics{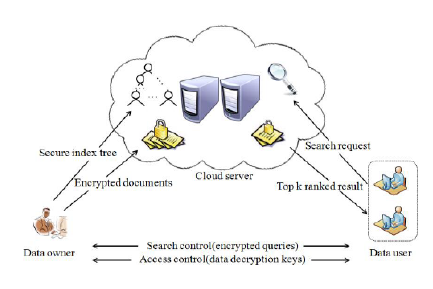}
    \caption{Framework of the search over encrypted cloud data.}
    \label{fig:fig14}
\end{figure}

\subsection{Cloud Search}
Our second module is Cloud Search and for this module, we also chose a comprehensive article for reference. However, this part of our project is the module which requires the least research effort in comparison to other project modules. Therefore, we have one referenced paper for the cloud module and only give an overview of the paper without mentioning any details of the approach.

\paragraph{Paper Review: Achieving Efficient Cloud Search Services: Multi-Keyword Ranked Search over Encrypted Cloud Data Supporting Parallel Computing~\cite{fu2015achieving}}

\paragraph{Overview: }
Nowadays, cloud computing has become quite popular. A large amount of data outsourced to the cloud by data owners for the purpose of accessing the large-scale computing resources and economic savings. \\

For data protection, the sensitive data should be encrypted by the data owner before outsourcing and this leads to the fact that traditional and efficient plain-text keyword search technique be-
comes useless. Therefore, researchers investigate how to design an efficient and effective searchable encryption scheme on cloud. You can see the overall framework in Figure~\ref{fig:fig14}.

\subsection{Query Expansion \& Suggestion}
~\emph{Lexical chasm} between queries and document titles is the main obstacle for improving base relevance of a search engine. This is substantially rooted from two things: i) authors of queries and documents are different (e.g. diverse vocabulary usage), ii) insufficient knowledge of technical terms in the corresponding domain. On the light of these, we consider to alleviate the~\emph{semantic gap} problem in the~\emph{Query Expansion} sub-module to improve the retrieval performance even more, in addition to the core methods utilized in our first module, LTR.
\subsubsection{Query Expansion}
In this sub-module, we aim to expand a given query with a selected set of terms to increase the coverage of the query. In selecting the terms, we can use three proposed methods which utilize three different word embeddings, word2vec~\cite{mikolov2013efficient}, FastText~\cite{joulin2016bag}, and GloVe~\cite{pennington2014glove}. We can combine the outputs of these three methods by ensemble learning (e.g. majority voting), for instance.

\paragraph{Paper Review: Query Expansion Using Word Embeddings~\cite{kuzi2016query}}

\paragraph{Motivation: }
Query expansion may help on improving retrieval performance of a search engine. For this, suitable terms should be selected to employ query expansion in a proper way, otherwise, it may deteriorate the retrieval performance by including irrelevant documents in returning document set.

\paragraph{Method: }
In this paper, researchers propose two term scoring methods for term selection. The main idea is to choose terms that are semantically related to the query by using word2vec's cbow em-
bedding approach applied over the entire search corpus. After computing scores and candidate terms are determined, the v terms assigned the highest score by a method M are used for query
expansion. Brief description of these M scoring methods are as follows.

\begin{itemize}
\item \underline{The centroid method:}
In this approach, researchers leverage the observation of adding
word2vec vectors representing terms that constitute an expression often yields a vector that semantically corresponds to the expression. Therefore, selecting procedure is employed by comparing cosine-similarity of a specific term's word2vec score in the collection and the corresponding query as a whole.

\item \underline{Fusion-based methods:}
Differently from the Cent method, in this approach, for each query term, $q_i$, a list $L{q_i}$ of its n most similar terms
t in the corpus according to cos($q_i$; t). Then,
cosine-similarity scores are used to compute softmax-normalized probabilities. After that, the resulting term lists are fused using Comb-SUM, CombMNZ and CombMAX~\cite{fox1994combination}.
\end{itemize}

\paragraph{Experimental Results: }
In evaluation, several TREC datasets are used as benchmark datasets. Results show statistically significant results, thus
the proposed idea can be used to improve the retrieval performance of a search engine by integrating also with other word embedding methods. \\

\paragraph{Paper Review: Using Word Embeddings
for Automatic Query Expansion~\cite{roy2016using}}

\paragraph{Motivation: }
The main goal of this work is again to find suitable terms to expand a query, i.e., increasing the coverage of the query without
including irrelevant terms.

\paragraph{Method: }
For selecting terms, researchers devise a query expansion technique, where related terms to a query are picked using K-nearest neighbour approach. More specifically, in the paper,
three kNN-based term selection methods are proposed as pre-retrieval kNN, post-retrieval kNN, pre-retrieval incremental kNN approaches. For an elaborate explanation of these methods, please refer to the paper.

\paragraph{Experimental Results: }
In the experimental results, the proposed method is evaluated on the standard ad-hoc task using both TREC collection and TREC web collection. The results indicate that, as an expansion method, the incremental method is generally safe; it produces performance
improvements for most of the queries and for only a few queries, it affects the performance badly. Therefore, this method can be integrated to our query expansion scheme, as well.\\

\paragraph{Paper Review: Query Expansion with
Locally-Trained Word Embeddings~\cite{diaz2016query}}

\paragraph{Motivation: }
Commonly used word embeddings, word2vec and GloVe, when trained globally underperform corpus and query-specific embeddings in the context of query expansion and retrieval tasks. Hence, in this paper researchers investigate the effect of local embeddings on query expansion.

\paragraph{Method: }
In order to generate local word embeddings, learning embeddings on topically-constrained corpora, instead of large topically-unconstrained corpora to utilize domain-specific
information. For this purpose, in this work, a language modelling approach is adopted to produce a query-specific set of topical documents.

\paragraph{Experimental Results: }
In the evaluation part, TREC datasets and ClueWeb 2009 corpora
are used and the evaluatin metric is chosen as NDCG@$10$. The comparison is done with the baseline method of query-likelihood and results show that the proposed query expansion approach
by using local word embeddings outperform the baseline on benchmark datasets. Based on this, in our query expansion scheme, we may give a try to local word embeddings, if feasible.\\

\subsubsection{Query Suggestion}
The main goal of this submodule is to improve user experience, i.e., enabling users to reach their searching information in a more accurate and quicker way. In the context of query suggestion,
there are two papers that can be utilized in our searching platform. These two articles~\cite{liu2017query, mei2008query} differ only in small details, therefore we will only mention about the first paper that contains the core idea.

\paragraph{Query Suggestion Using Hitting Time~\cite{mei2008query}}

\paragraph{Motivation: }
This paper aims to generate query suggestions while ensuring semantic consistency with the original query not to lose context
information.

\paragraph{Method: }
In this work, finding query suggestion procedure is mainly as follows. Initially, a bipartite graph is established on click-logs and it is used for query expansion. Then on the graph, hitting time from a given query to other queries are modelled by random walking for finding candidate query suggestions.

\paragraph{Experimental Results: }
Results show that this technique is beneficial for query suggestion and can be incorporated to our platform to improve user search experience. We selected this work for query suggestion, because of two reasons: i) it is one of the most comprehensive and successful method in this area, and ii) for efficiency purposes, since we already construct a bipartite-graph on click-through logs and we can use the same graph also for query suggestion.\\

\section{Patents}
\subsection{US20160004776 A1, 2016~\cite{cloud2016}}
In this patent, cascading searching is used to locate a desired person in a social media ecosystem. The social media search system is provided on cloud and we utilize this patent to scale our
search service on a cloud platform.
\subsection{US7689520B2, 2010~\cite{burges2010machine}}
This patent proposes a machine learning system to rank a set of documents by using differentiable parameters. The ranking will be optimized according to a cost module that computes cost scores on a pair of examples. This patent helps on generating letor-based ranking models to improve our system.

\bibliography{main}
\bibliographystyle{plain}

\end{document}